\documentclass{article}
\usepackage{subfig}
\usepackage{spconf,amsmath,epsfig}
\usepackage{flushend, cuted}
\usepackage{amsmath}
\usepackage{multicol}
\usepackage{amssymb}
\usepackage{booktabs} 
\usepackage{url}
\usepackage{multirow}
\usepackage{graphicx}
\usepackage{enumerate}
\usepackage{bm}
\usepackage{graphicx}
\usepackage{color}
\usepackage{gensymb,textcomp}
\usepackage[linesnumbered,ruled]{algorithm2e}
\let\OLDthebibliography\thebibliography
\renewcommand\thebibliography[1]{
	\OLDthebibliography{#1}
	\setlength{\parskip}{0pt}
	\setlength{\itemsep}{0pt plus 0.3ex}
}

\graphicspath{
	{./images/}
}
\begin{document}\sloppy
	
	\def\x{{\mathbf x}}
	\def\L{{\cal L}}

	\title{Spatial Attention-based Non-reference Perceptual Quality Prediction Network for   Omnidirectional Images}

	%
	\name{Li Yang, Mai Xu, Deng Xin and Bo Feng}
	\address{School of Electronic and Information Engineering, Beihang University, Beijing 100191, China}

	\maketitle

	\begin{abstract}
		
		Due to the strong correlation between visual attention and perceptual quality, many methods attempt to use human saliency information for image quality assessment. Although this mechanism can get good performance, the networks require human saliency labels, which is not easily accessible for omnidirectional images (ODI). To alleviate this issue, we propose a spatial attention-based perceptual quality prediction network for non-reference quality assessment on ODIs (SAP-net). To drive our SAP-net, we establish a large-scale IQA dataset of ODIs (IQA-ODI), which is composed of subjective scores of 200 subjects on 1,080 ODIs. In IQA-ODI, there are 120 high quality ODIs as reference, and 960 ODIs with impairments in both JPEG compression and map projection. 
		Without any human saliency labels, our network can adaptively estimate human perceptual quality on impaired ODIs through a self-attention manner, which significantly promotes the prediction performance of quality scores. Moreover, our method greatly reduces the computational complexity in quality assessment task on ODIs. Extensive experiments validate that our network outperforms 9 state-of-the-art methods for quality assessment on ODIs. The dataset and code have been available on \url{ https://github.com/yanglixiaoshen/SAP-Net}.    	
		
	\end{abstract}
	\vspace{-1.1em}
	\section{Introduction}
	\label{sec:intro}
	\vspace{-0.9em}
	With the rapid development of virtual reality (VR), omnidirectional images (ODIs), as a new type of multimedia, have played an increasingly important role in human life. Different from 2D images, ODIs offer an interactive and immersive visual experience with high resolution. The ultra-high resolution of ODIs poses great challenges on current image processing
	systems, e.g., streaming \cite{Hosseini2016Adaptive}, compression \cite{xu2020state} and transmission \cite{otsuka2006transpost}, etc. Generally, ODIs are projected to planes and compressed heavily, which dramatically degrades the quality of experience (QoE). Therefore, it is crucial to research the ODI image quality assessment (IQA), to guide the
	optimization of image processing systems.
	
	In the past decades, many works have emerged for exploring IQA on 2D images, which can be classified into three categories: full-reference IQA (FR-IQA) \cite{wang2004image, zhang2011fsim, wang2010information, zhang2014vsi, sheikh2006image, liu2011image}, reduced reference IQA (RR-IQA) \cite{rehman2012reduced} and non-reference IQA (NR-IQA) \cite{min2016blind, moorthy2011blind, zhang2015feature, mittal2012making, mittal2012no}. Specifically, the peak signal-to-noise ratio (PSNR) and structural similarity index (SSIM) are the most representative methods for FR-IQA, which measure the pixel-wise and structure-wise distortion between the reference and distorted image, respectively. Moreover, due to the absence of high quality reference, most NR-IQA methods evaluate image degradation based on natural scene statistic (NSS) features, via wavelet, DCT and Laplace transform. Recently, the great success of convolutional neural network (CNN) \cite{liu2020mrs, he2016deep, RBQE, Guan_2021} boosts the IQA performance significantly \cite{Kang2014Convolutional, bosse2017deep, ma2017dipiq, kim2017deep}. 
	Although many works have been proposed for FR/NR IQA on 2D images, there still exists limited research on FR/NR IQA for ODIs or omnidirectional videos (ODVs) \cite{sun2019mc360iqa, xu2020viewport, li2018bridge, truong2019non}. Specifically, in \cite{truong2019non}, patch sampling and quality score pooling strategies are proposed for NR-IQA based on ODI equator-bias technique. Considering the projection distortion of ODI, MC360IQA \cite{sun2019mc360iqa} develops a viewport-based multi-channel CNN for NR-IQA, via projecting ODI into six equal cube faces. In \cite{li2018bridge} and \cite{xu2020viewport}, the head movement and eye movement of human behavior are incorporated into CNN to weight the quality scores of compressed ODVs. Although the above methods take human behavior or equator-bias characteristic of ODIs/ODVs into account for IQA, the pre-processing tasks, i.e., viewport prediction, patch sampling, pay much more computational expense compared with the main task of IQA. Besides, the viewport extraction \cite{sun2019mc360iqa} and patch sampling \cite{truong2019non} strategies are both designed in hand-crafted manner, which may not hold for all distortion types. Thus, these issues make the IQA performance unsatisfactory  and lead to large computational complexity.           
	
	\begin{figure*}[!tb]
		\begin{center}
			\includegraphics[width=1\linewidth]{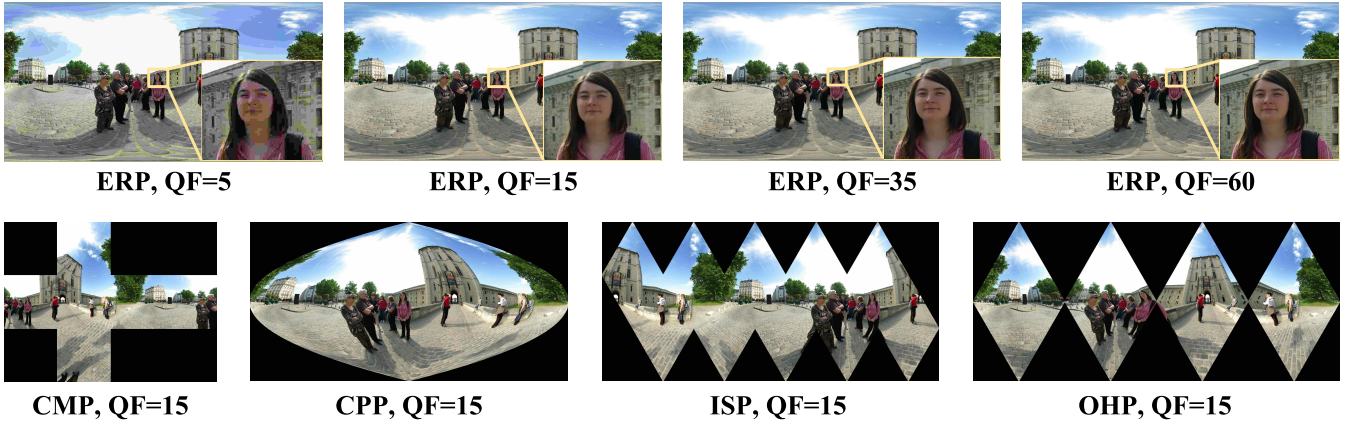}
		\end{center}
		\vspace{-2em}
		\caption{\small Examples of different map projection types and JPEG compression levels in our IQA-ODI dataset.}
		\vspace{-1.4em}
		\label{dataset}
	\end{figure*}
	
	To tackle the above issues, we propose a spatial-attention based perceptual quality prediction network (SAP-net) for NR-IQA on ODIs \footnote{Contact me (Email: 13021041@buaa.edu.cn) for the dataset and code}. To drive our SAP-net, we establish a large-scale IQA dataset of ODIs (IQA-ODI), which is composed of subjective scores of 200 subjects on 1,080 ODIs. In IQA-ODI, there are 120 high quality ODIs as reference, and 960 ODIs with impairments in both JPEG compression and map projection. Then, we mine our dataset for some data analyses on subjective scores. Based on the data analysis, we develop three novel modules in our SAP-net, including wavelet-based residual enhancement (WBRE) module, perceptual quality estimation (PQE) module and quality regression (QR) module. Specifically, the impaired ODI is fed into WBRE for an objective quality enhancement. Then, as a pseudo reference, the enhanced ODI is exploited to generate the visual error map, which indicates the objective degradation between the pseudo reference and impaired ODI. Subsequently, the error map and impaired ODI are incorporated into PQE, to predict the perceptual quality map. At last, the perceptual quality map which implies the human visual sensitivity is input to QR for the regression of the final quality score. The experimental results show that our SAP-net significantly advances the state-of-the-art performance of objective NR-IQA on ODIs. 
	
	\vspace{-0.1em}
	\begin{table}[!t]%
		\centering%
		\vspace{-1.2em}
		\caption{\footnotesize The IQA performance of pixel-wise and structure-wise methods.}
		\label{tab:anal1}%
		\scriptsize
		\begin{tabular}{ccccccc}%
			\hline
			\multicolumn{1}{|c|}{\multirow{2}{*}{Metrics}} & \multicolumn{3}{c|}{Pixel-wise IQA} & \multicolumn{3}{c|}{Structure-wise IQA} \\
			\cline{2-7} \multicolumn{1}{|c|}{} & PSNR & S-PSNR & \multicolumn{1}{c|}{CPP-PSNR} & SSIM & FSIM & \multicolumn{1}{c|}{IWSSIM}\\
			\hline
			\multicolumn{1}{|c|}{PLCC} &0.485 & 0.542 & \multicolumn{1}{c|}{0.512} & 0.571 & 0.925 & \multicolumn{1}{c|}{0.883}    \\ 
			\multicolumn{1}{|c|}{SROCC} & 0.397 &0.429 & \multicolumn{1}{c|}{0.401} & 0.518 & 0.886 & \multicolumn{1}{c|}{0.880} \\ 
			\multicolumn{1}{|c|}{MAE} &9.491 &8.987 & \multicolumn{1}{c|}{9.222}  & 8.743 & 3.926 & \multicolumn{1}{c|}{4.938}  \\  \hline	
		\end{tabular}%
		\vspace{-2em}
	\end{table}%
	\vspace{-1em}
	\section{Dataset and data analysis}
	\label{IQA_dataset}
	\vspace{-0.9em}
	\subsection{Dataset establishment}
	\textbf{Stimuli.}
	Our IQA-ODI dataset has in total 1,080 ODIs, of which 120 are reference with a wide range of content categories, such as human, landscape and nature. The resolutions of all references are 8K (7680 $\times$ 3840 pixels) under equirectangular projection (ERP) format which can ensure the high quality. Then, two kinds of impairment are taken into account: compression level and projection pattern. Specifically, the former is measured by different quality factors, introduced by the common image compression standard, JPEG \cite{wallace1992jpeg}. The latter is a unique characteristic of ODIs. In all, we consider four quality factors ($q^f$) = 5, 15, 35 and 60, and 4 projection patterns: CMP, CPP, ISP and OHP \cite{xu2020state} for impairment on reference. On account of the general use in map projection on ODIs, we choose ERP as the main format in impaired ODIs. Accordingly, we build two modes for obtaining the impaired ODIs. For mode 1, the projection pattern is set to ERP and $q^f$ is set to 5, 15, 35 and 60, respectively. For mode 2, $q^f$ is set to 15 and the projection pattern is set to CMP, CPP, ISP and OHP, respectively. Some examples of the two modes can be seen in Figure \ref{dataset}.
	
	\textbf{Procedure.}
	The total number of subjects participating in our experiment is 200, consisting of 138 males and 62 females. The age of subjects ranges from 17 to 33. All the subjects are divided into 10 groups to view 108 ODIs (12 reference and 96 impaired ODIs each group), such that each subject only watches one group of ODIs for avoiding eye fatigue. For the experiment, we use HTC Vive as a media palyer, which is connected to a high-performance computer. The viewing time for each ODI is set to 20 seconds, and there is a 5-minute interval in the middle of total viewing process to avoid eye fatigue and motion sickness. During the experiments, the subjects wearing the HTC Vive are asked to sit in a comfortable swivel chair, which enables them to rotate 360$^\circ$ freely. As such, all panoramic regions in the ODI can be easily accessed. As the viewing time of each stimuli ends, subject is asked to give a subjective score according to the quality of stimuli. Then, the score is transmitted to the computer for data processing.  
	\begin{table}[!t]%
		\centering%
		\vspace{-1.2em}
		\caption{\footnotesize The IQA performance of FR- and NR-IQA methods.}
		\label{tab:anal2}%
		\scriptsize
		\begin{tabular}{ccccccc}%
			\hline
			\multicolumn{1}{|c|}{\multirow{2}{*}{Metrics}} & \multicolumn{3}{c|}{FR-IQA} & \multicolumn{3}{c|}{NR-IQA} \\
			\cline{2-7} \multicolumn{1}{|c|}{} & VSI & VIF & \multicolumn{1}{c|}{GSM} & DIVINE & ILNIQE & \multicolumn{1}{c|}{NIQE}\\
			\hline
			\multicolumn{1}{|c|}{PLCC} &0.816 & 0.851 & \multicolumn{1}{c|}{0.896} & 0.222 & 0.648 & \multicolumn{1}{c|}{0.740}    \\ 
			\multicolumn{1}{|c|}{SROCC} & 0.886 &0.741 & \multicolumn{1}{c|}{0.898} & 0.182 & 0.488 & \multicolumn{1}{c|}{0.698} \\ 
			\multicolumn{1}{|c|}{MAE} &6.190 &5.492 & \multicolumn{1}{c|}{4.617}  & 9.845 & 8.042 & \multicolumn{1}{c|}{6.609}  \\  \hline	
		\end{tabular}%
		\vspace{-1em}
	\end{table}%

	\begin{figure}[!tb]
		\begin{center}
			\includegraphics[width=1\linewidth]{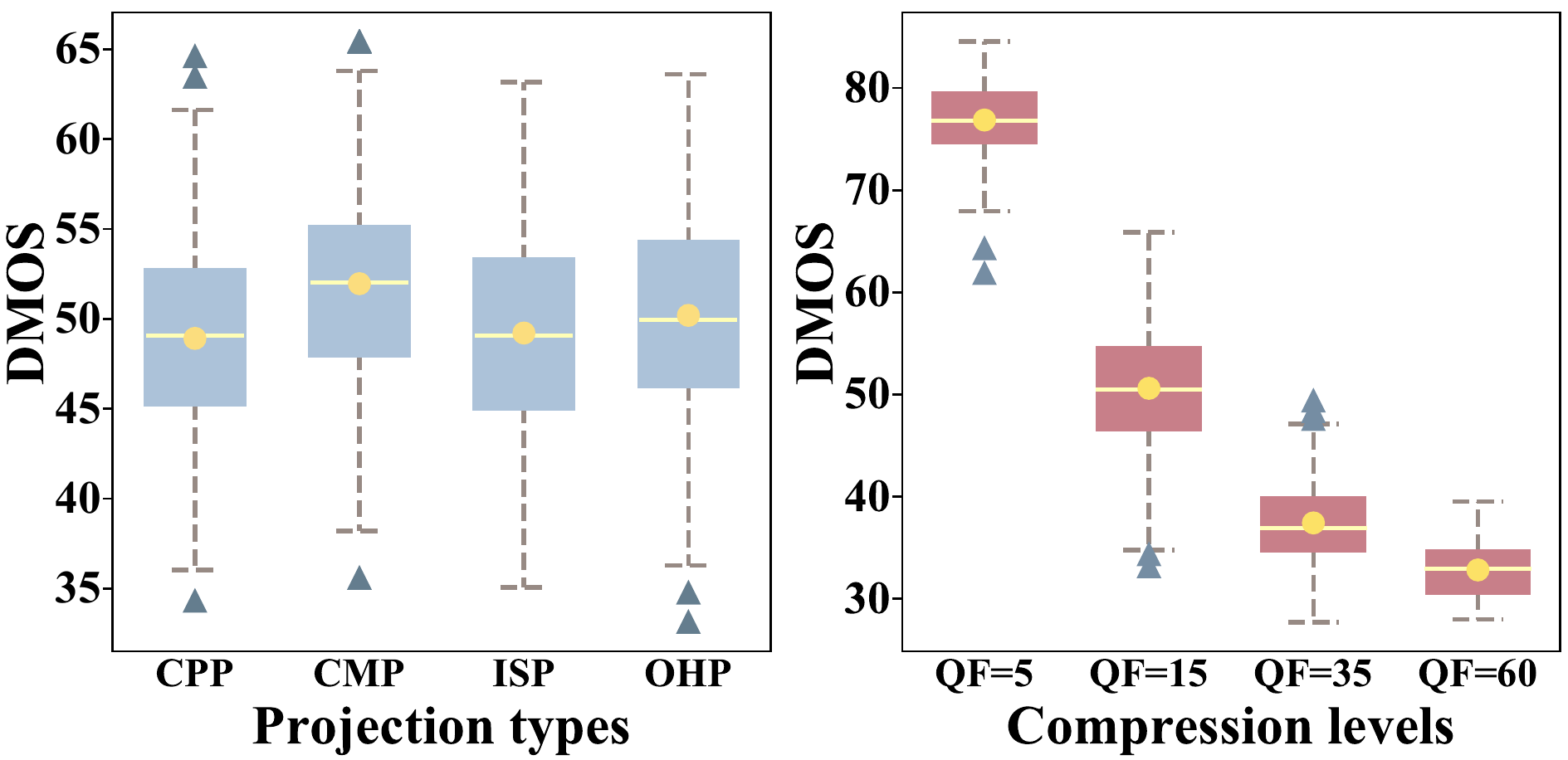}
		\end{center}
		\vspace{-1.8em}
		\caption{\small DMOS under different projections and compression levels.}
		\vspace{-1.7em}
		\label{boxplot}
	\end{figure}
	\begin{figure*}[!tb]
		\begin{center}
			\includegraphics[width=1\linewidth]{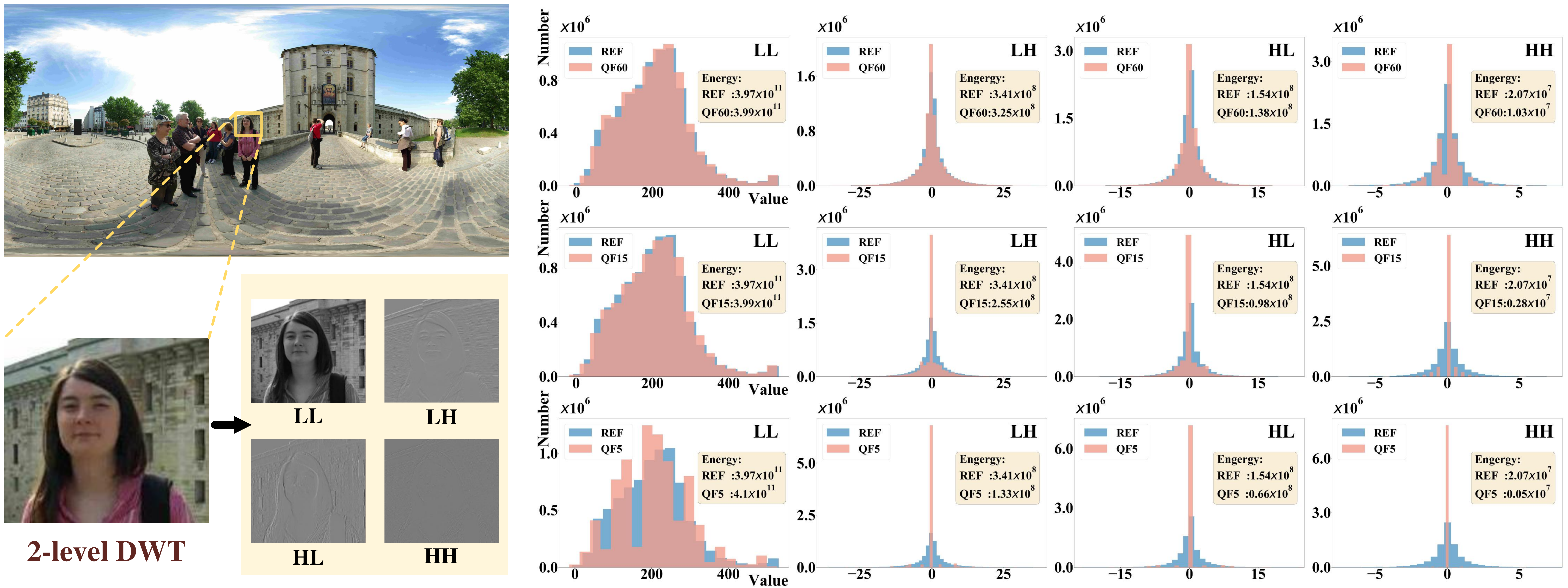}
		\end{center}
		\vspace{-1.6em}
		\caption{\small Left: An example of DWT on an impaired ODI; Right: The histogram of wavelet energies on LL, LH, HL and HH sub-bands, across different compression levels as well as the corresponding reference. }
		\vspace{-1.6em}
		\label{dwt}
	\end{figure*}
	\textbf{Data processing.} 
	After obtaining all the subjective scores of all ODIs, we focus on the data processing. First, we calculate the mean opinion score (MOS) of each ODI by averaging across 20 subjective scores. Then, we compute difference MOS between the impaired ODI and the corresponding reference, to discount any subject preference for certain reference. Following the method \cite{li2018bridge}, we can obtain the differential mean opinion score (DMOS) which indicates the subjective quality measurement for each ODI. Note that the MOS of each reference has actual value, while its DMOS value is always 0. All the ODIs DMOS lie in the range of [0, 100]. The higher the DMOS is, the worse quality the ODI possesses.   
	
	\subsection{Dataset analysis}
	\vspace{-0.8em}
	\textbf{Analysis on DMOS.}
	First, we focus on the subjective DMOS at different impairment types in our IQA-ODI dataset. Figure \ref{boxplot} shows the boxplots of DMOS under different projection types and compression levels, respectively. From this figure, we can find that the number of outliers are few which indicates the DMOS over all impaired ODIs are normal and reliable. 
	Additionally, we find that the median lines in the four boxes are close to each other in the left boxplot (value 49.1, 52.0, 49.1 and 49.9), while that of the right one is far (value 76.8, 50.5, 36.9 and 32.9). It indicates that compression levels have much more impact on subjective quality of impaired ODIs, when compared with projection types.

	Next, we evaluate the performance of different objective IQA methods over all impaired ODIs, by calculating the correlation between the objective IQA scores and the corresponding DMOS. 
	Here, the correlation is evaluated by spearman's rank correlation coefficient (SROCC), pearson correlation coefficient (PLCC), and mean absolute error (MAE), the results of which can be seen in Table \ref{tab:anal1} and Table \ref{tab:anal2}. Specifically, Table \ref{tab:anal1} shows the evaluation performance of pixel-wise IQA methods (including PSNR, S-PSNR \cite{Yu2015A} and CPP-PSNR \cite{zakharchenko2016quality}) and structure-wise IQA methods (including SSIM \cite{wang2004image}, FSIM \cite{zhang2011fsim} and IWSSIM \cite{wang2010information}). Table \ref{tab:anal2} shows the evaluation performance of traditional FR-IQA methods (including VSI \cite{zhang2014vsi}, VIF \cite{sheikh2006image}, GSM \cite{liu2011image}) and NR-IQA methods (including DIVINE \cite{moorthy2011blind}, ILNIQE \cite{zhang2015feature}, NIQE \cite{mittal2012making}). From Table \ref{tab:anal1}, we can find that the structure-wise IQA methods performs better than the pixel-wise IQA methods. This indicates that the structure-wise IQA methods correlate well with subjective perceived quality on impaired ODIs. Moreover, Table \ref{tab:anal2} implies that the NR-IQA method performs much worse than the FR-IQA methods, since the lack of high quality ODI as reference makes the NR-IQA performance degrades.    
	
	\begin{table}[!t]%
		\centering%
		\vspace{-1.em}
		\caption{\small The average energy loss of all impaired ODIs.}
		\footnotesize
		\label{tab:anal3}%
		\begin{tabular}{|c|cccc|}%
			\hline
			Sub-bands & QF = 5 & QF = 15 & QF = 35 & QF = 60 \\
			\hline
			LL & 0.52\% & 0.24\% & 0.04\% & 0.02\% \\
			LH & 44.19\% & 29.01\% & 7.32\% & 3.52\% \\
			HL & 51.75\% & 38.06\% & 11.91\% & 5.03\%  \\
			HH & 90.05\% & 70.79\% & 49.21\% & 29.89\%  \\
			\hline		
		\end{tabular}%
		\vspace{-1.9em}
	\end{table}%

	\textbf{Wavelet analysis on ODIs.}
	Inspired by \cite{reisenhofer2018haar, wang2020multi, deng2019wavelet}, we deploy two-level haar discrete wavelet transform (DWT) on each impaired ODI and its corresponding reference, resulting in four sub-bands called LL, LH, HL and HH. Here, LL is the low-frequency sub-band, LH, HL and HH are the
	sub-bands with high-frequency information at horizontal, vertical and diagonal directions, respectively. Figure \ref{dwt} (left) shows an example of DWT processing on an impaired ODI. Then, in Figure \ref{dwt} (right), we plot the histograms of wavelet coefficients (also called energies) of LL, LH, HL and HH sub-bands, across different compression levels as well as the corresponding reference. The histogram figure illustrates that the low-frequency sub-band LL of each compression level has quite similar histogram distributions as the reference, but that is not the case for high-frequency sub-bands of LH, HL and HH. Moreover, it is obvious that as the compression level increases, the energy of the high-frequency sub-bands declines more severely. Table \ref{tab:anal3} tablets the average loss of energies for each sub-band across all impaired ODIs, when compared with the corresponding reference. It shows the similar results as the above analyses. The correlation between compression levels and wavelet sub-bands energies motivates us to embed the DWT for impaired ODI enhancement in our SAP-net. 
	\begin{figure*}[!tb]
		\begin{center}
			\includegraphics[width=1\linewidth]{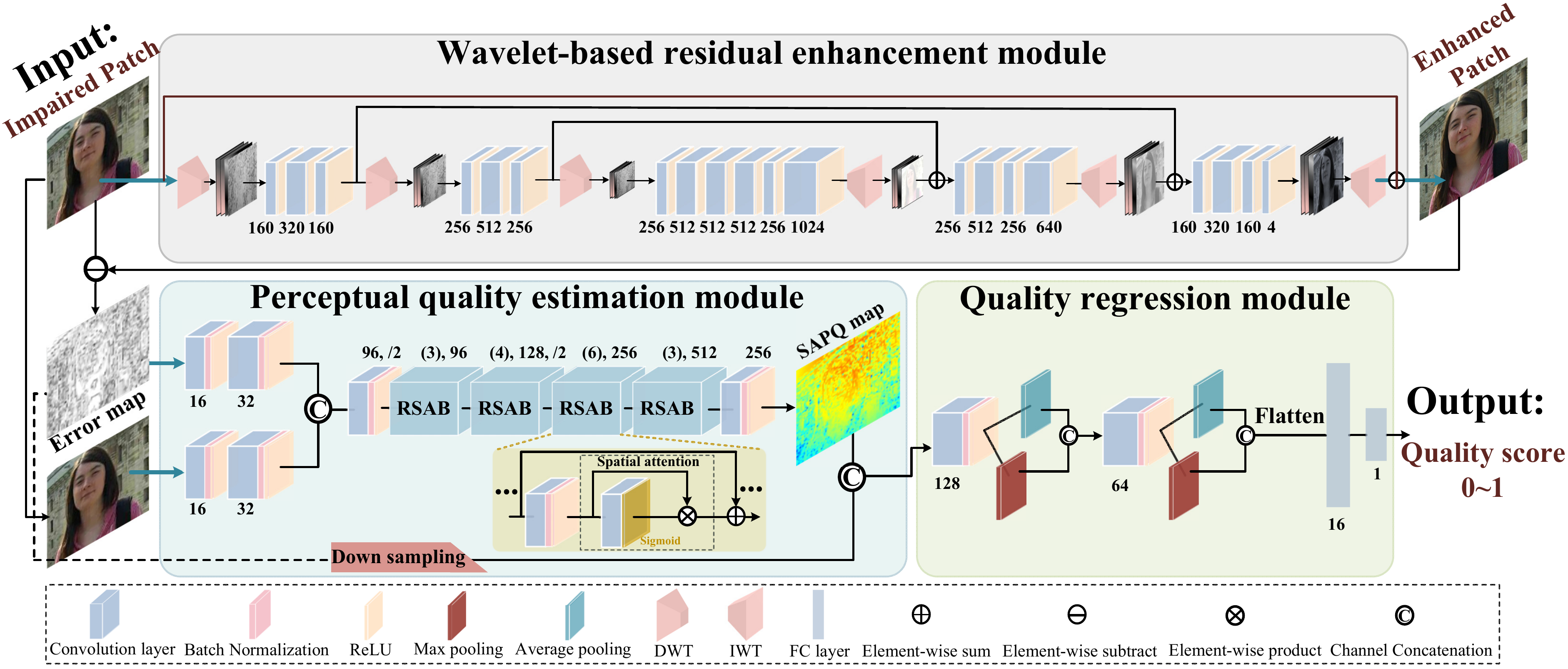}
		\end{center}
		\vspace{-1.8em}
		\caption{\footnotesize The framework of our SAP-net for NR-IQA. For the PQE module, we use ResNet-34 as backbone, which has 3, 4, 6 and 3 residual blocks, respectively.}
		\vspace{-1.6em}
		\label{framework}
	\end{figure*}
	
	\vspace{-1.2em}
	\section{The proposed SAP-net}
	\label{sap-net}
	\vspace{-0.8em}
	The architecture of the proposed SAP-net is shown in Figure \ref{framework}, which contains three novel modules, i.e., WBRE, PQE and QR.
	Given an input patch $\mathbf{I}$ of impaired ODI, the ultimate goal of SAP-net is to predict the NR-IQA score $\hat{s}$ for approximating the ground-truth DMOS $s$. In the following, we introduce the specific function of each module for facilitating the prediction of NR-IQA score.  
	
	\textbf{WBRE module.} 
	To take advantage of the reference in FR-IQA, we adopt the image enhancement network to generate a pseudo reference $\mathbf{\hat{R}}$ by following the mechanism of ``residual in residual" in \cite{zhang2018image}. To obtain better performance, we further make some modifications. First, the pooling operation at each level of WBRE is properly replaced by DWT for down-sampling. This down-sampling operation can avoid information loss cause by pooling, and benefit the quality enhancement. Meanwhile, it efficiently enlarges the receptive field to improve the capacity of feature
	representation. Second, we extend the ``residual in residual" structure to multi-level features in the WBRE, which can be seen in Figure \ref{framework}. In addition to the long skip connection between input $\mathbf{I}$ and output $\mathbf{\hat{R}}$, we generally combine the DWT and inverse discrete wavelet transform (IWT) features from different levels, via long or short skip connections. It ensures that abundant low-frequency information can be bypassed, and benefits the high-frequency feature enhancement of detailed texture.  
	Note that the final outputs of WBRE are the wavelet sub-bands $\mathbf{\hat{F}} = \{\mathbf{\hat{F}_{LL}}, \mathbf{\hat{F}_{LH}}, \mathbf{\hat{F}_{HL}}, \mathbf{\hat{F}_{HH}}\}$, as well as the enhanced patch $\mathbf{\hat{R}}$ which is obtained by combining $\mathbf{\hat{F}}$ via IWT.
	
	\textbf{PQE module.} After the WBRE, the pseudo reference $\mathbf{\hat{R}}$ is used to generate the visual error map $\mathbf{\hat{E}}$ by element-wise subtracting with the impaired patch $\mathbf{I}$. This visual error map indicates the objective degradation between $\mathbf{\hat{R}}$ and $\mathbf{I}$. Then, both the error map $\mathbf{\hat{E}}$ and impaired patch $\mathbf{\hat{R}}$ are incorporated in a complementary manner into the PQE for further perceptual quality estimation. Specifically, $\mathbf{\hat{E}}$ and $\mathbf{\hat{R}}$ are fed into different convolutional layers at the beginning, and concatenated after the second convolutional layer. Subsequently, the concatenated feature flows into convolutional layers and several residual spatial attention block (RSAB). To consider the strong correlation between the visual attention and perceptual quality \cite{zhang2014vsi}, we integrate spatial attention mechanism into a residual block \cite{he2016deep} in each RSAB for attention-based quality estimation. Not like \cite{li2018bridge} needs to take the human saliency map as one part of input to compute the quality map, our RSAB can implicitly learn the visual attention on the impaired patch via the self-attention mechanism. With no need for human saliency labels to supervise, it significantly alleviates the computation complexity. Moreover, RSAB makes full use of the interaction of the error map and impaired patch, where the error map can be a guider to promote the estimation of visual attention. At last, we can obtain the spatial attention-based perceptual quality (SAPQ) map $\mathbf{\hat{P}}$ for further processing. Especially, $\mathbf{\hat{P}}$ is employed as a weighting mask for the error map, to reflect the fusion of attention- and quality-importance of a local region. Not like the element-wise production operation between  the quality map and error map \cite{kim2017deep}, we use the channel concatenation to combine $\mathbf{\hat{P}}$ and $\mathbf{\hat{E}}$ as the input feature of the following QR, denoted as $\mathbf{\hat{C}}$.  
	
	\textbf{QR module.} The main task of QR is to regress the predicted quality feature $\mathbf{\hat{C}}$ into a quality score $\hat{s}$. Here, we design a simple yet efficient structure, which is composed of two-layer CNN, max-pooling, average-pooling and two fully-connected (FC) layers. Note that we utilize both the features filtered from max-pooling and average-pooling, to increase the robustness of quality score regression. Finally, the objective NR-IQA score $\hat{s}$ of the impaired patch $\mathbf{I}$ can be obtained, and the quality score of the whole impaired ODI can be obtained by averaging the scores of all patches. 
	
	\begin{table*}[!t]%
		\centering%
		\small
		\caption{\small Comparison on IQA performance between our and other methods, over all test impaired ODIs.}
		\label{tab:exp1}%
		\begin{tabular}{ccccccccccc}%
			\hline
			\multicolumn{1}{|c|}{\multirow{2}{*}{Metrics}} & \multicolumn{3}{c|}{FR-IQA methods} & \multicolumn{6}{c|}{NR-IQA methods} & \multicolumn{1}{c|}{Our method} \\
			\cline{2-11} \multicolumn{1}{|c|}{} & S-PSNR & FSIM & \multicolumn{1}{c|}{GSM} & BRISQUE & PSS & CNNIQA  & WaDIQaM-NR  & MC360IQA &\multicolumn{1}{c|}{dipIQ} & \multicolumn{1}{c|}{SAP-net}  \\
			\hline
			\multicolumn{1}{|c|}{PLCC $\uparrow$} &0.4810 & 0.7865 & \multicolumn{1}{c|}{0.9014} & 0.8770 & 0.8996 & 0.8994 & 0.6443 & 0.7017 &\multicolumn{1}{c|}{0.8741} & \multicolumn{1}{c|}{$\textbf{0.9258}$}    \\ 
			\hline
			\multicolumn{1}{|c|}{SROCC $\uparrow$} & 0.3580 &0.8889 & \multicolumn{1}{c|}{0.9006} & 0.8171 & 0.8174 & 0.7395 & 0.5176 &  0.6660 &\multicolumn{1}{c|}{0.7213} &\multicolumn{1}{c|}{$\textbf{0.9036}$} \\
			\hline 
			\multicolumn{1}{|c|}{KROCC $\uparrow$} & 0.2247 & 0.7200 & \multicolumn{1}{c|}{0.7368} & 0.6315 & 0.6299 & 0.5535 & 0.3629 &0.4842 &\multicolumn{1}{c|}{0.5282} &\multicolumn{1}{c|}{$\textbf{0.7396}$} \\
			\hline 
			\multicolumn{1}{|c|}{RMSE $\downarrow$} & 11.9731 & 8.4345 & \multicolumn{1}{c|}{5.9123} & 6.5617 & 5.9639 & 5.9700 & 10.4433 &9.7293 &\multicolumn{1}{c|}{6.6338} &\multicolumn{1}{c|}{$\textbf{4.7845}$} \\
			\hline 
			\multicolumn{1}{|c|}{MAE $\downarrow$} & 10.1265 & 6.7686 & \multicolumn{1}{c|}{4.7173} & 5.0586 & 4.7885 & 4.6373 & 8.5610 &7.3182 &\multicolumn{1}{c|}{5.3513} &\multicolumn{1}{c|}{$\textbf{3.6078}$} \\
			\hline 	
		\end{tabular}%
		\vspace{-0.8em}
	\end{table*}%
	\begin{figure*}[!tb]
		\begin{center}
			\includegraphics[width=1\linewidth]{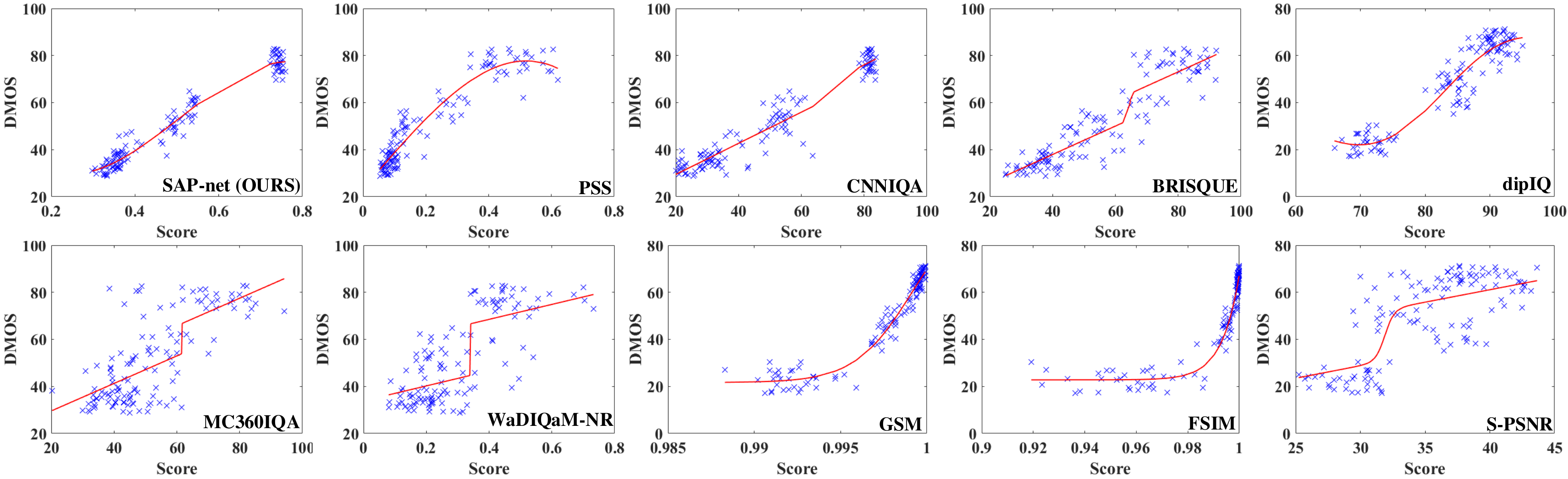}
		\end{center}
		\vspace{-2.2em}
		\caption{\footnotesize The scatter plots of the objective IQA scores versus the DMOS values over all impaired ODIs in the test set. The logistic fitting curves are also shown.}
		\vspace{-1.4em}
		\label{subjective_exp1}
	\end{figure*}
	\begin{figure}[!t]%
		\centering%
		\subfloat[Ablating the RSAB in PQE module.]{\label{fig:ab:cm}%
			\includegraphics[width=0.86\linewidth]{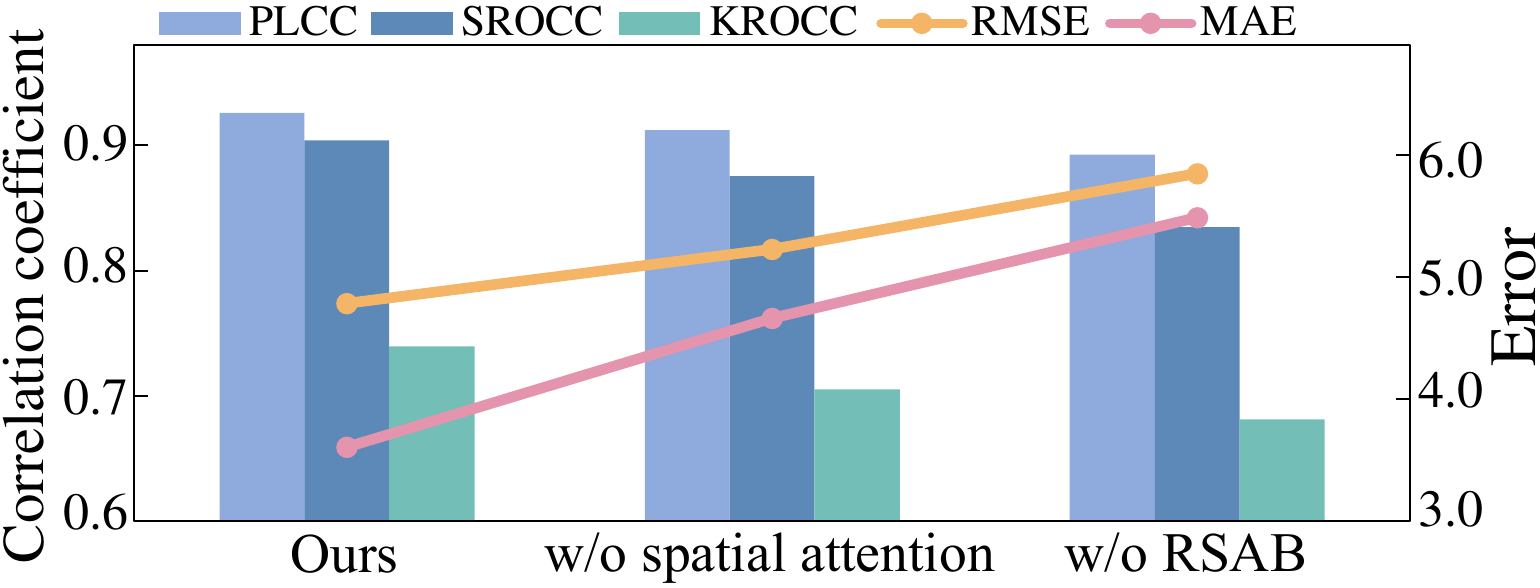}%
		}\hfil%
		\vspace{-0.1em}
		\subfloat[Ablating the map concatenation settings.]{\label{fig:ab:hm}%
			\includegraphics[width=0.86\linewidth]{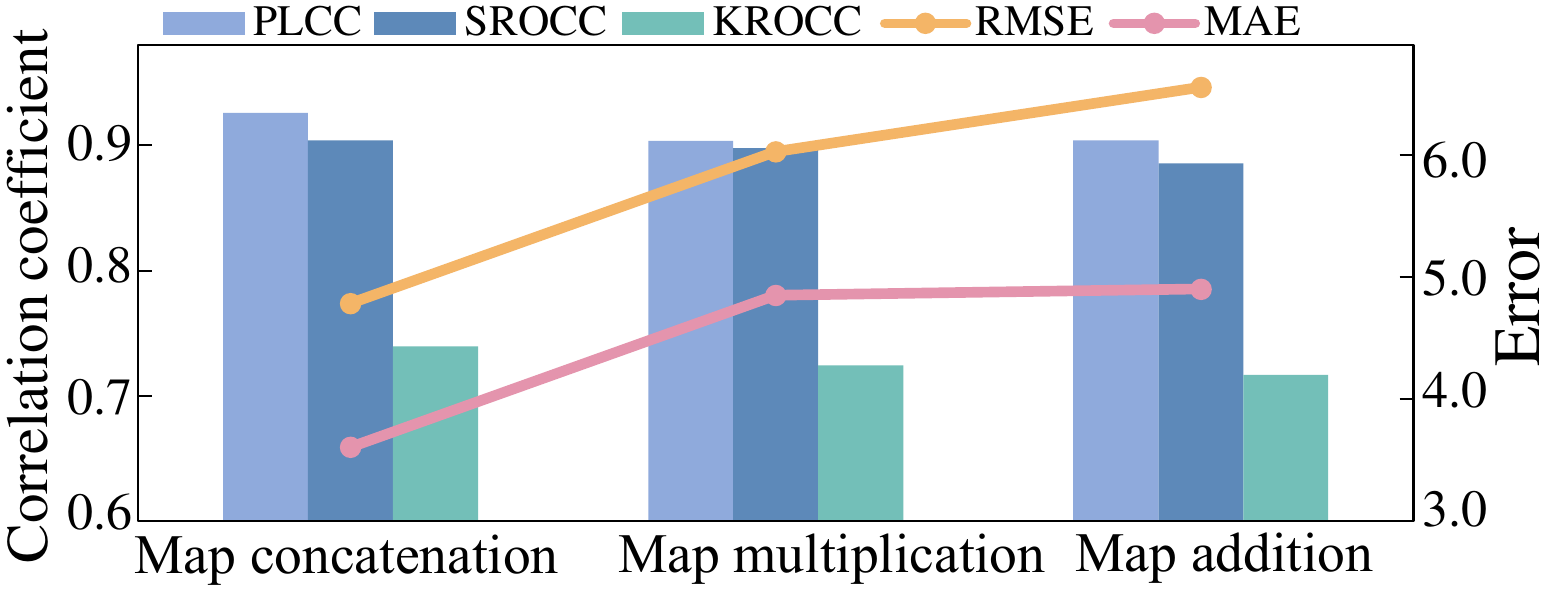}%
		}\hfil%
		\vspace{-0.8em}
		\caption{Results of ablation experiments of the SAP-net.}\label{fig:ab}%
		\vspace{-1.7em}
		\label{ablaz}
	\end{figure}%
	
	\textbf{Loss function. }
	Here, we discuss about the loss function $\mathcal L$ for training our SAP-net, which is formulated as
	\vspace{-0.7em}
	\begin{equation}
	\label{loss}
	\begin{aligned}
	\mathcal L = \mathcal L_e + \lambda_1 \mathcal L_a,
	\end{aligned}
	\vspace{-0.6em}
	\end{equation} 
	where $\mathcal L_e$ denotes the quality enhancement loss for training the WBRE module, and $\mathcal L_a$ denotes the quality assessment loss for training the PQE and QR module. Note that $\lambda_1$ is a hyper-parameter to balance $\mathcal L_e$ and $\mathcal L_a$.
	Specifically, the loss $\mathcal L_e$ models how close the predicted wavelet sub-bands $\mathbf{\hat{F}}$ are to the ground truth $\mathbf{F}$. It is defined by a weighted Charbonnier penalty function \cite{lai2017deep} (a differentiable variant of $\mathcal \ell_1$ norm) in wavelet domain as follows:
	\vspace{-0.6em}
	\begin{equation}
	\label{loss_m}
	\begin{aligned}
	\mathcal L_e = \sqrt{{\bm{\beta}}^{1/2} \odot ||\mathbf{\hat{F}} - \mathbf{F}||_c^2 + \epsilon_{\beta}} ,
	\end{aligned}
	\vspace{-0.6em}
	\end{equation} 
	where  $\mathbf{F}=\{\mathbf{{F}_{LL}}, \mathbf{{F}_{LH}}, \mathbf{{F}_{HL}}, \mathbf{{F}_{HH}}\}$ are the ground truth wavelet sub-bands, and $\odot$ represents dot product. Moreover, $\bm{\beta}$ is the weight matrix to balance the importance of each sub-band and $\epsilon_{\beta}$ is a scaling parameter set to 0.001. For the supervision of PQE and QR module, the goal is to minimize the MSE loss between the predicted score $\hat{s}$ and the ground truth DMOS $s$, which is formulated by 
	\vspace{-0.6em}
	\begin{equation}
	\label{loss_m}
	\begin{aligned}
	\mathcal L_a = ||\hat{s} - s||_2^2 .
	\end{aligned}
	\vspace{-0.6em}
	\end{equation} 
	Finally, with the target of loss minimization, the parameters of our SAP-net are updated using the stochastic gradient descent algorithm with the Adam optimizer.
	\vspace{-1.4em}
	\section{Experimental results}
	\label{exp}
	\vspace{-1em}
	\textbf{Settings.}
	In our experiments, 960 impaired ODIs in our ODI-IQA dataset are randomly divided into training and testing sets in a ratio of 5 : 1, i.e., 800 training and 160 test impaired ODIs. During training stage, the input impaired ODI is divided into patches with a fixed size 256$\times$256, and the patches are randomly extracted from ODI region with non-overlapping at different iterations. The learning rate for training our SAP-net is initially set to 1$\times$10$^{-4}$, and the batch size is set to be 8. The hyper-parameter $\lambda_1$ is set to be 10.
	
	\textbf{Performance evaluation.}
	For performance evaluation, the correlation between the DMOS values and objective scores of each
	IQA method is measured on 160 test ODIs, in terms of PLCC, SROCC, Kendall rank-order correlation coefficient (KROCC), Root Mean Squared Error (RMSE) and MAE. Note that the larger values of PLCC, SROCC and KROCC indicate higher correlation, while the smaller
	values of RMSE and MAE mean higher prediction accuracy. 
	Here, we follow \cite{sun2019mc360iqa} to employ a logistic function for fitting the objective IQA scores to their corresponding DMOS values, such that the fitted scores of all IQA methods have the same scale as DMOS. Then, we compare the performance of our SAP-net with 9 state-of-the-art methods on NR-IQA or FR-IQA, including S-PSNR, FSIM, GSM, BRISQUE \cite{mittal2012no}, PSS \cite{min2016blind}, CNNIQA \cite{Kang2014Convolutional}, WaDIQaM-NR \cite{bosse2017deep}, MC360IQA \cite{sun2019mc360iqa} and dipIQ \cite{ma2017dipiq}. Note that CNNIQA, WaDIQaM-NR, MC360IQA and dipIQ are all trained over our training set.
	Table \ref{tab:exp1} tabulates the comparison of quantitative IQA results. From it, we can find that our SAP-net significantly outperforms all other NR-IQA methods, with at least 0.026, 0.086, 0.108, 1.179 and 1.030 improvements in PLCC, SROCC, KROCC, RMSE and MAE, respectively. This verifies the effectiveness of the designs in our method.
	Moreover, it can be obviously seen from Figure \ref{subjective_exp1} that the IQA scores of our method have much higher correlation with the DMOS values, compared
	with all others. Thus, we can conclude that the SAP-net performs much better than other methods.
	
	\textbf{Ablation experiments.}
	Here, we analysis the impact of RSAB and map concatenation settings on NR-IQA performance, and conduct two ablation experiments respectively: (1) Ablation on the RSAB and spatial attention scheme in PQE module; (2) Ablation on map concatenation between error map and SAPQ map. Figure \ref{ablaz} shows that the settings in our SAP-net can achieve the best ablation results when compared with all other settings.

	\vspace{-1.2em}
	\section{Conclusion}
	\label{conc}
	\vspace{-1.em}
	In this paper, for NR-IQA on ODIs, we have established a large-scale dataset containing 1,080 ODIs with five projection types and four compression levels. Furthermore, we proposed a SAP-net which emphasizes the incorporation of attention and perceptual quality for NR-IQA on ODIs. Our SAP-net achieved superior effectiveness and efficiency compared with other baselines. In the future, we may consider the unsupervised manner to devise our network for NR-IQA on ODIs.


	\vspace{-1.1em}
	\bibliographystyle{IEEEbib}
	\bibliography{icme2021template}
	
\end{document}